\documentclass{pazhastl}
\usepackage{graphicx}
\usepackage{natbib}

\def\plotspectrum#1#2#3{
  \begin{minipage}{1.0\columnwidth}
    \begin{minipage}{0.049\columnwidth}
      \rotatebox{90}{\phantom{0000}#3}
    \end{minipage}
    \begin{minipage}{0.9\columnwidth}
      \includegraphics[bb=58 188 556 497,width=0.97\columnwidth]{#1}
      \centerline{#2}
    \end{minipage}
    
    \vskip 3pt
    ~
  \end{minipage}
}

\def\smfigurewocap#1#2#3{
  \begin{minipage}{1.0\columnwidth}
    \begin{minipage}{0.049\columnwidth}
      \rotatebox{90}{\phantom{0000}#3}
    \end{minipage}
    \begin{minipage}{0.9\columnwidth}
      \includegraphics[bb=58 188 556 678,width=0.97\columnwidth]{#1}
      \centerline{#2}
    \end{minipage}
    
    \vskip 3pt
    ~
  \end{minipage}
}

\def\oiii{\mbox{[OIII]}}

\begin{document}

\journalinfo{2008}{34}{0}{0}[0] 

\title{Several New Active Galactic Nuclei Among X-ray Sources Detected by
  \emph{INTEGRAL} and \emph{SWIFT} Observatories}

\author{
  R.~A.~Burenin\address{1}\email{rodion@hea.iki.rssi.ru},
  A.~V.~Mescheryakov\address{1},
  M.~G.~Revnivtsev\address{1,2},
  S.~Yu.~Sazonov\address{1,2},
  I.~F.~Bikmaev\address{3,4},
  M.~N.~Pavlinsky\address{1},
  R.~A.~Sunyaev\address{1,2}
  \addresstext{1}{Space Research Institute (IKI), ul. Profsoyuznaya 84/32,
    Moscow, Russia}
  \addresstext{2}{Max-Planck-Institut f\"ur Astrophysik,
    Karl-Schwarzschild-Strasse 1, D-85740 Garching bei M\"unchen, Germany}
  \addresstext{3}{Kazan State University, ul. Kremlevskaya 18, Kazan,
    Russia}
  \addresstext{4}{Academy of Sciences of Tatarstan, ul. Baumana, 20, Kazan,
  Russia}
}

\shortauthor{Burenin et al.} 
\shorttitle{New AGNs among \emph{INTEGRAL} and \emph{SWIFT} X-ray sources}
\submitted{December 26, 2007}

\begin{abstract}

  We present the results of the optical identifications of a set of X-ray
  sources from the all-sky surveys of \emph{INTEGRAL} and \emph{SWIFT}
  observatories.  Optical data were obtained with Russian-Turkish 1.5-m
  Telescope (RTT150).  Nine X-ray sources were identified as active galactic
  nuclei (AGNs). Two of them are hosted by nearby, nearly exactly edge-on,
  spiral galaxies MCG~$-$01$-$05$-$047 and NGC\,973. One source,
  IGR\,J16562$-$3301, is most probably BL Lac object (blazar). Other AGNs
  are observed as stellar-like nuclei of spiral galaxies, with broad emission
  lines in their spectra.

  For the majority of our hard X-ray selected AGNs, their hard X-ray
  luminosities are well-correlated with the luminosities in \oiii,5007
  optical emission line. However, the luminosities of some AGNs deviate from
  this correlation. The fraction of these objects can be as high as 20\%.
  In particular, the flux in \oiii\ line turns to be lower in two nearby
  edge-on spiral galaxies, which can be explained by the extinction in their
  galactic disks.


\end{abstract}

\section*{Introduction}

The all-sky surveys in hard X-rays give a valuable information on the
populations of objects, which are difficult to observe due to the presence
of absorption in optical and soft X-rays. The most sensitive to date are
recently completed \emph{RXTE} \citep[XSS,][]{mikej04} and \emph{INTEGRAL}
\citep{krivonos07} all-sky surveys, and on-going \emph{SWIFT} all-sky survey
\citep{tueller07}.

These surveys contains a large number of known nearby active galactic nuclei
(AGNs), cataclysmic variables, X-ray binaries, located in Galactic plane.
However, it turns out, that a considerable number of sources, up to 20--30\%
of their total number, can not be identified with previously known objects.
In order to make a complete use of the capabilities of these surveys for
statistical studies, one need to identify all the X-ray sources with optical
objects and to determine their nature.

The work on optical identifications of hard X-ray sources from the all-sky
surveys was started on Russian-Turkish 1.5-m Telescope (RTT150)
\citep{bikmaev2006a,bikmaev2006b}. In this paper we present the results on
the extragalactic objects --- active galactic nuclei, obtained in frames of
this work during Autumn 2006 -- Spring 2007. Some of these results were
immediately published in astronomical circulars
\citep{atel880,atel883,atel948,atel1270}. Here we discuss the results of
these observations in more detail.

\section{Observations} 
\label{sec:obs}

A characteristic property of X-ray telescopes on board \emph{INTEGRAL} and
\emph{SWIFT} observatories is that their accuracy of the localization of
X-ray sources, in many cases, is insufficient to unambiguously associate the
X-ray source with a certain object in optical band. The error box of
\emph{INTEGRAL} and \emph{SWIFT} hard X-ray sources usually is about few
arcminutes. When observing in optical band, even at high Galactic latitudes,
one will find a considerable number of galaxies and stars in the error box
of that size. At low Galactic latitudes, the field of that size contain so
large number of stars, that it is impossible to directly associate the X-ray
source with any optical object.

In order to obtain more accurate positions of hard X-ray sources one can use
the observations in softer X-ray band, where the positions of X-ray sources
can be measured much more accurately. First, one can use the data of
\emph{ROSAT} All-Sky Survey \citep{voges99}. In case of strongly absorbed
X-ray sources, which are not detected in \emph{ROSAT} All-Sky Survey, one
can use the data of pointed observation of \emph{Chandra}, \emph{XMM}-Newton
telescopes or X-ray telescope aboard \emph{SWIFT}, in harder X-ray band.

The data for some hard X-ray sources were obtained in dedicated Chandra
observations \citep{sazonov05,sazonov08}. An extensive program of
observations of \emph{SWIFT} and \emph{INTEGRAL} hard X-ray sources is also
carried out with \emph{SWIFT} X-ray telescope \citep[XRT, see,
e.g.,][]{tueller07}. We downloaded all necessary additional open data from
\emph{HEASARC}\footnote{http://heasarc.nasa.gov/} archive.

For our optical observations we first selected the hard X-ray sources, with
the positional accuracy enhanced in one or another way.  All optical data
were obtained with Russian-Turkish 1.5-m Telescope (RTT150), using medium
and low resolution spectrometer
\emph{TFOSC}\footnote{http://astroa.physics.metu.edu.tr/tug/tfosc.html}. In
most cases we used 54~mkm slit, which corresponds to 0.96\arcsec\ size in
the sky. We mostly used grism \#15, which give the best throughput and the
most wide spectral coverage (3500--9000~\AA). In this setup, the spectral
resolution was $\approx12$~\AA\ (FWHM). In those cases, where we needed
higher spectral resolution, we used other grisms. The data were reduces
using standard \emph{IRAF}\footnote{http://iraf.noao.edu} software.

\begin{table*}
  \centering
  \caption{Optical identifications of hard X-ray sources}
  \label{tab:src}
  \medskip
  \begin{tabular}{lccclcllccc}
    \hline
    \hline
    Name & $\alpha$ & $\delta$ & $m_R$ & Type & $z$\phantom{0$^{**}$} & Other\\
             & \multicolumn{2}{c}{(J2000)} & & &                         & name \\
    \hline
    IGR\,J01528$-$0326 & 01 52 48.90 & $-$03 26 47.9 & \phantom{1}9.3 & Sy2 & 0.017197$^{*}$ &   MCG~$-$01$-$05$-$047\\ 
    IGR\,J02343$+$3229 & 02 34 20.27 & $+$32 30 18.6 & \phantom{1}9.8 & Sy2 & 0.016195$^{*}$ &  NGC\,973\\  
    IGR\,J03334$+$3718 & 03 33 18.78 & $+$37 18 11.4 & 13.5 & Sy1 & 0.05583\phantom{0$^{*}$}  & \\  
    IGR\,J13038$+$5348 & 13 03 59.50 & $+$53 47 30.2 & 12.9 & Sy1 & 0.03016\phantom{0$^{*}$}  &  MCG~$+$09$-$21$-$096\\ 
    IGR\,J16562$-$3301 & 16 56 16.77  & $-$33 02 12.8 & 17.5 & BL Lac & ?\phantom{0$^{*}$} &  \\
    SWIFT\,J1745.4$+$2906 & 17 45 38.28 & $+$29 08 22.2 & 14.2 & Sy1 & 0.11040\phantom{0$^{*}$} &  \\ 
    SWIFT\,J1930.5$+$3414 & 19 30 13.81 & $+$34 10 49.8 & 15.8 & Sy1 & 0.06326\phantom{0$^{*}$} & \\ 
    1RXS\,J193347.6$+$325422 & 19 33 47.16 & $+$32 54 26.2 & 13.4 & Sy1 & 0.05794\phantom{0$^{*}$} & \\ 
    RX\,J2135.9$+$4728 & 21 35 54.02 & $+$47 28 22.3 & 13.2 & Sy1 & 0.02523\phantom{0$^{*}$}  & SWIFT\,J2156.2$+$4728 \\   
    \hline 
  \end{tabular}
  
  \medskip
  \begin{minipage}{1.0\linewidth}
    $^{*}$  Redshifts are taken from \emph{NED}: http://nedwww.ipac.caltech.edu/
  \end{minipage}
\end{table*}


\section{Results}
\label{sec:res}

The list of X-ray sources, which we were able to identify in Autumn 2006 --
Spring 2007, is presented in Table~\ref{tab:src}. In the Table for each
X-ray source we give its name, coordinates, R magnitude, type of the object
and redshift. The coordinates are given in 2000 epoch, using USNO B1.0
catalog \citep{monet03} as a reference frame. For nearby galaxies we give
the coordinates of active nuclei or the maximum of emission near the center
of the galaxy, where the optical spectrum was obtained. R band magnitudes
are given for reference and are taken from \emph{USNO-B1.0} catalog. The
designation of object types are as follows: Sy1,2 --- Seyfert galaxy of type
1 or 2; BL Lac --- BL Lacertae object. The AGNs redshifts were measured
using only narrow lines, like \oiii\ 4959,5007, [NI] 5199, [OI] 6300, [SII]
6716,6731, and also in some cases [OII] 3727, [NeIII] 3869 and narrow parts
of Balmer lines. The errors of $z$ measurement should not be larger than
$0.0001$.

\begin{figure*}
  \centering
  \begin{tabular}[c]{ll}
    \large IGR\,J01528$-$0326 & \large IGR\,J02343$+$3229 \\
    \includegraphics[width=0.8\columnwidth]{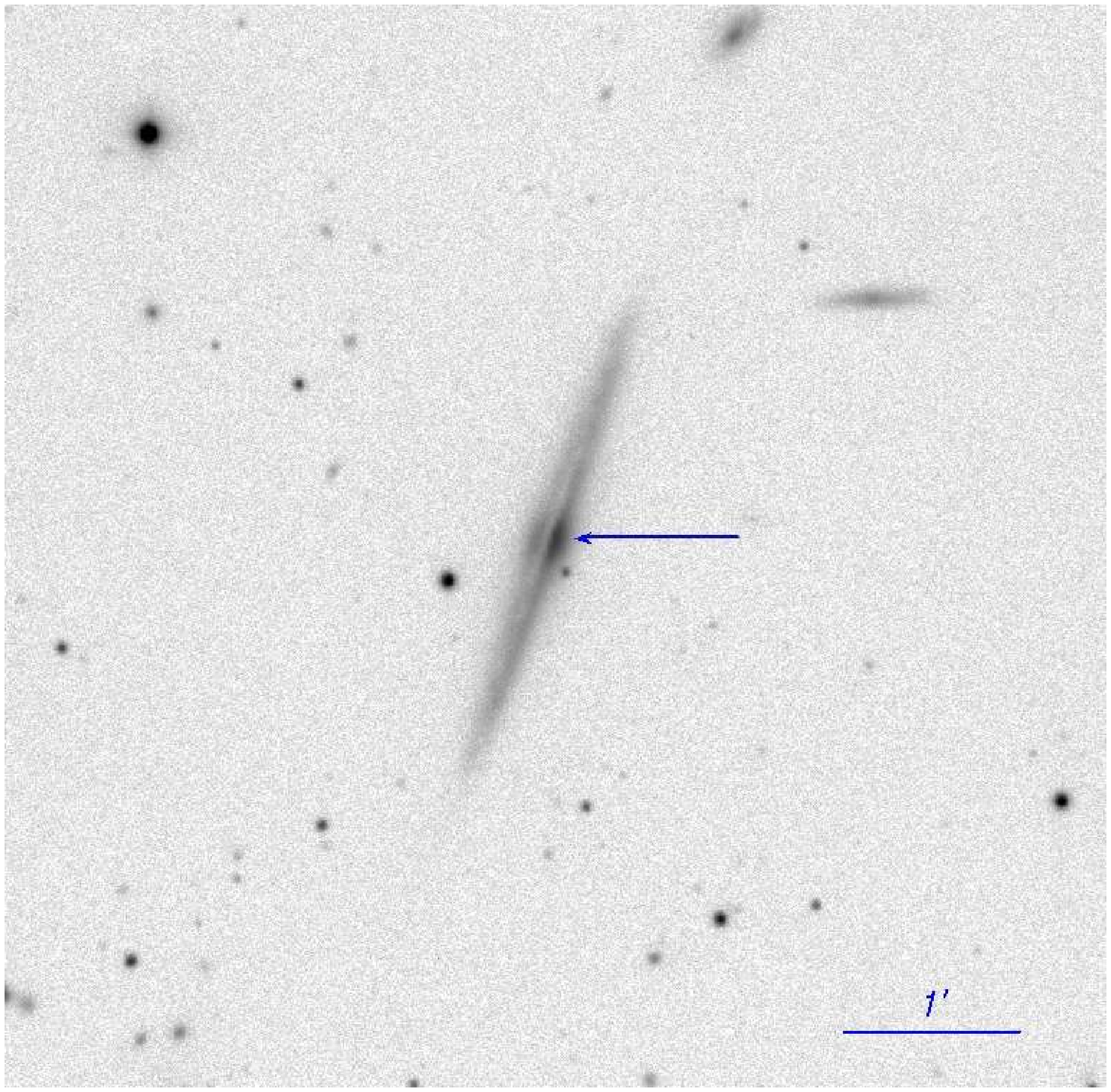} & 
    \includegraphics[width=0.8\columnwidth]{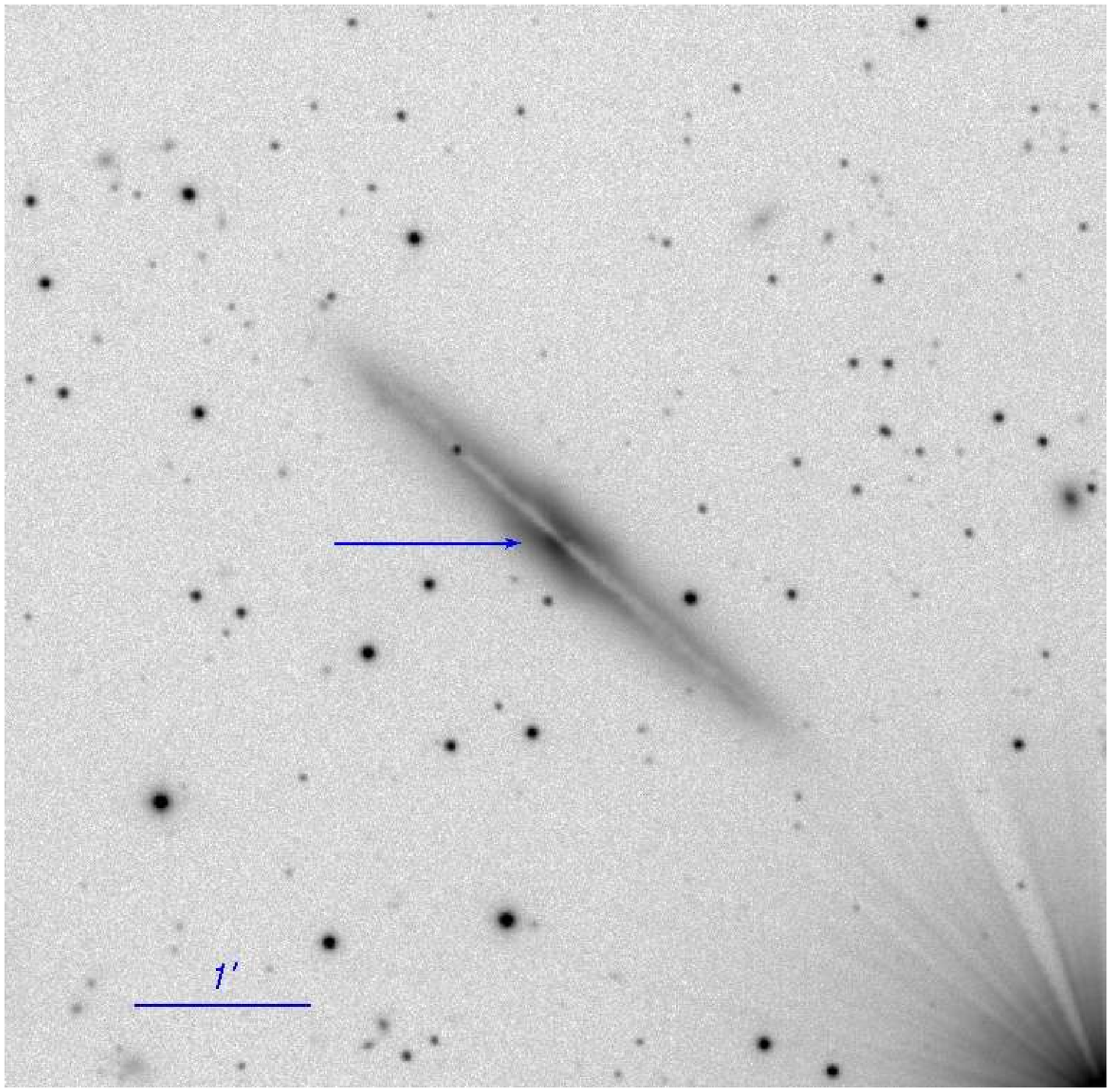} \\
    \\
    \large IGR\,J03334$+$3718 & \large IGR\,J13038$+$5348 \\
    \includegraphics[width=0.8\columnwidth]{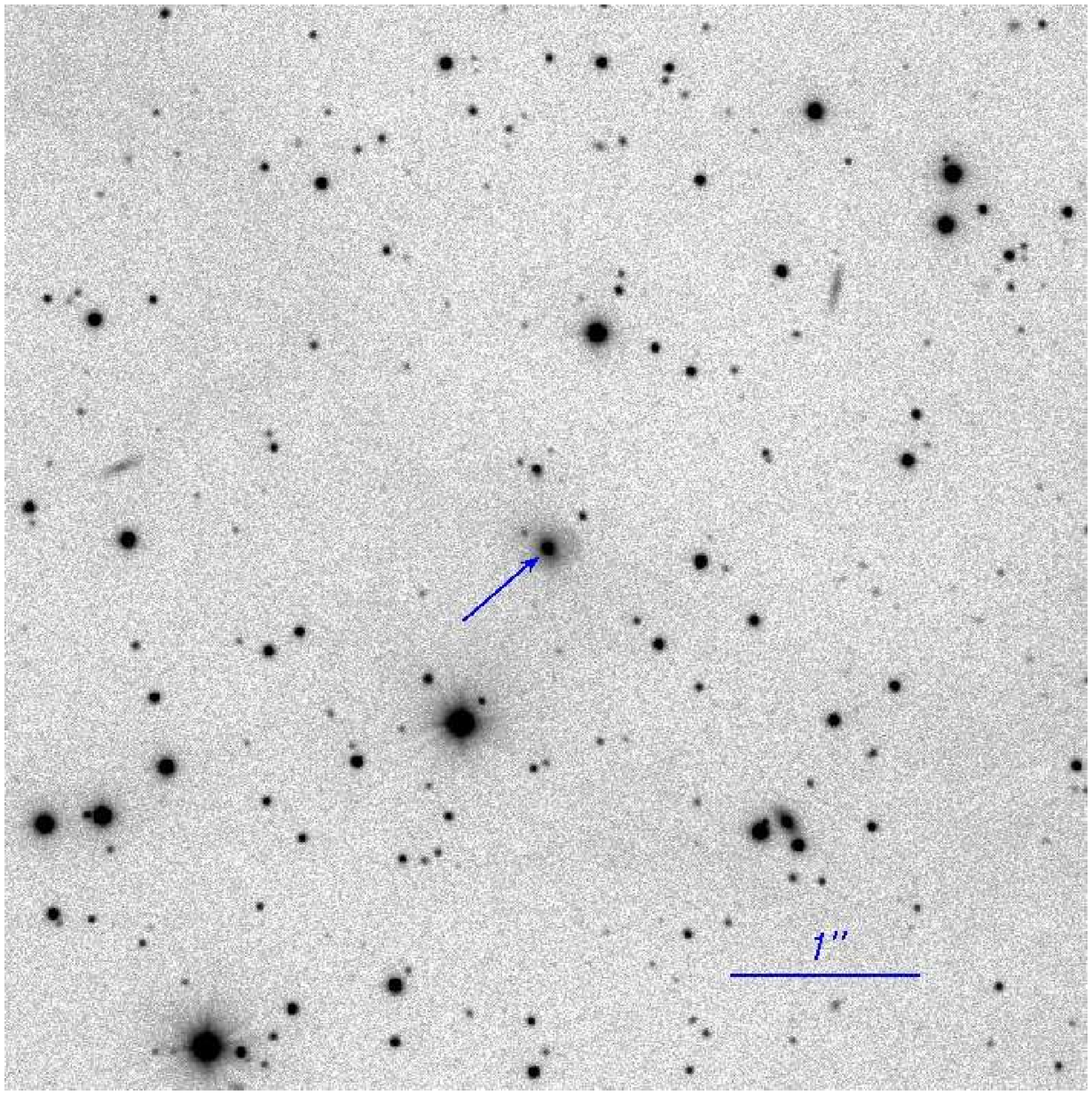} & 
    \includegraphics[width=0.8\columnwidth]{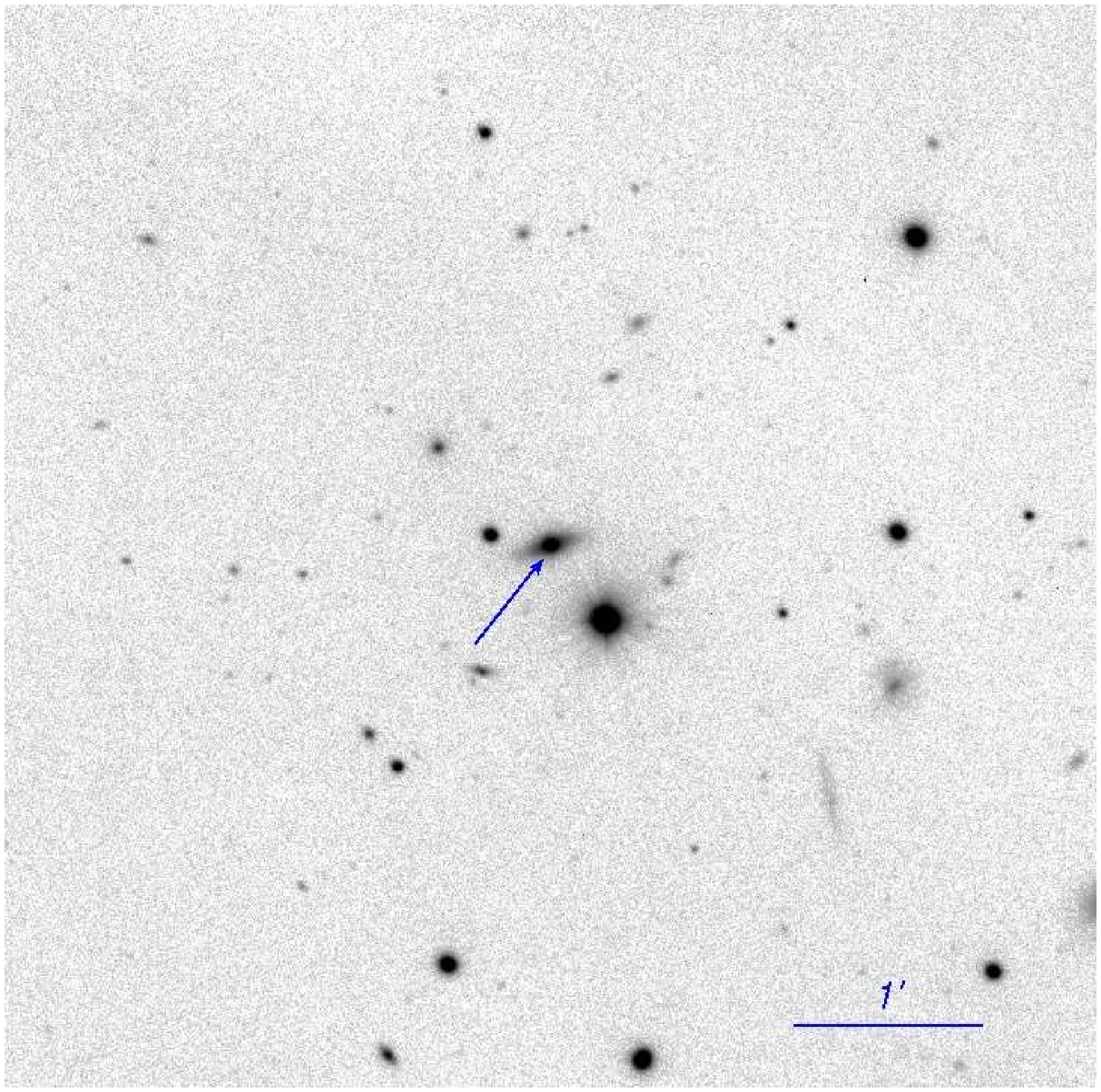} \\
    \\
    \large IGR\,J16562$-$3301, \emph{DSS-II-r} & \large SWIFT\,J1745.4$+$2906 \\
    \includegraphics[width=0.8\columnwidth]{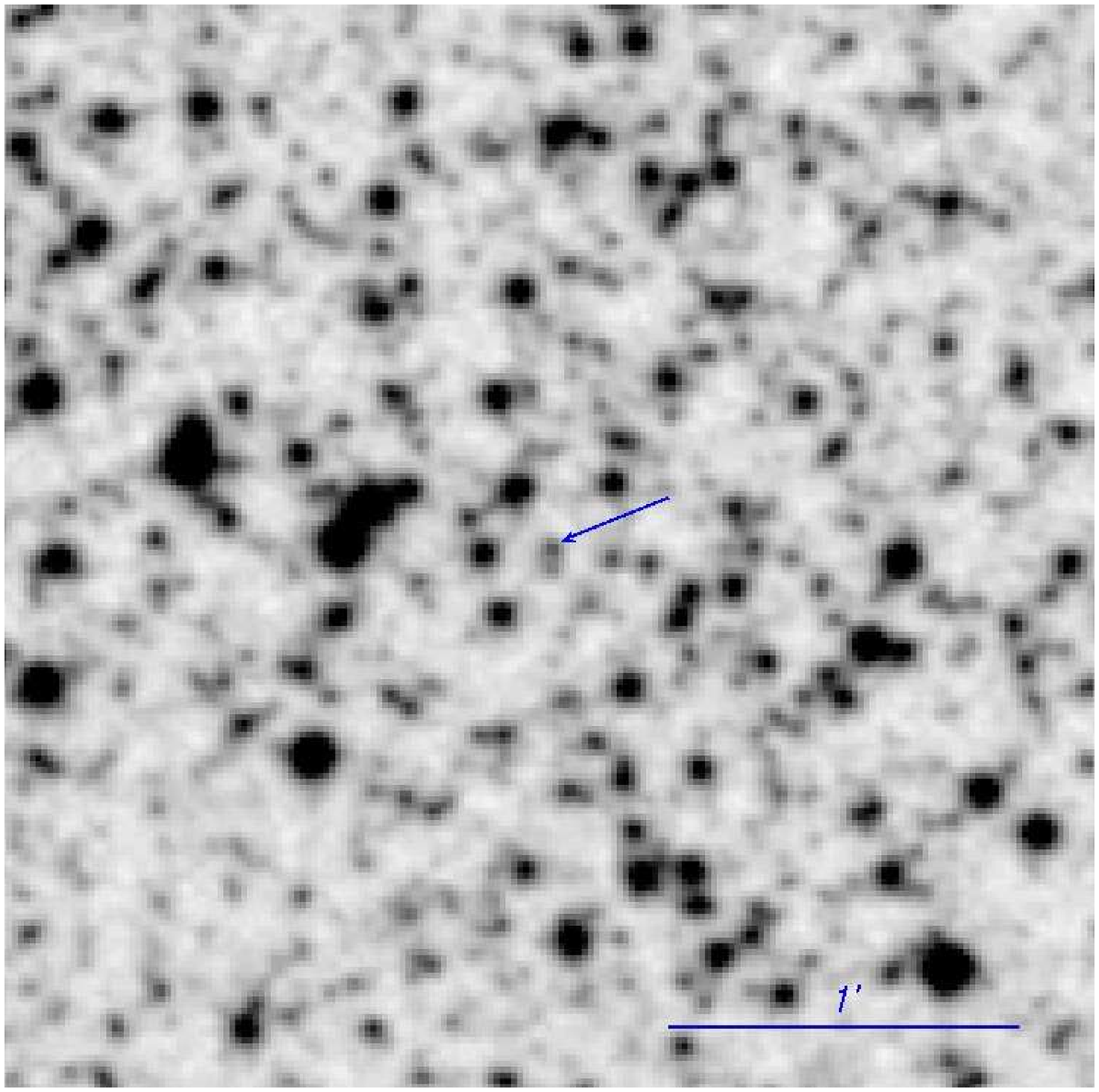} &
    \includegraphics[width=0.8\columnwidth]{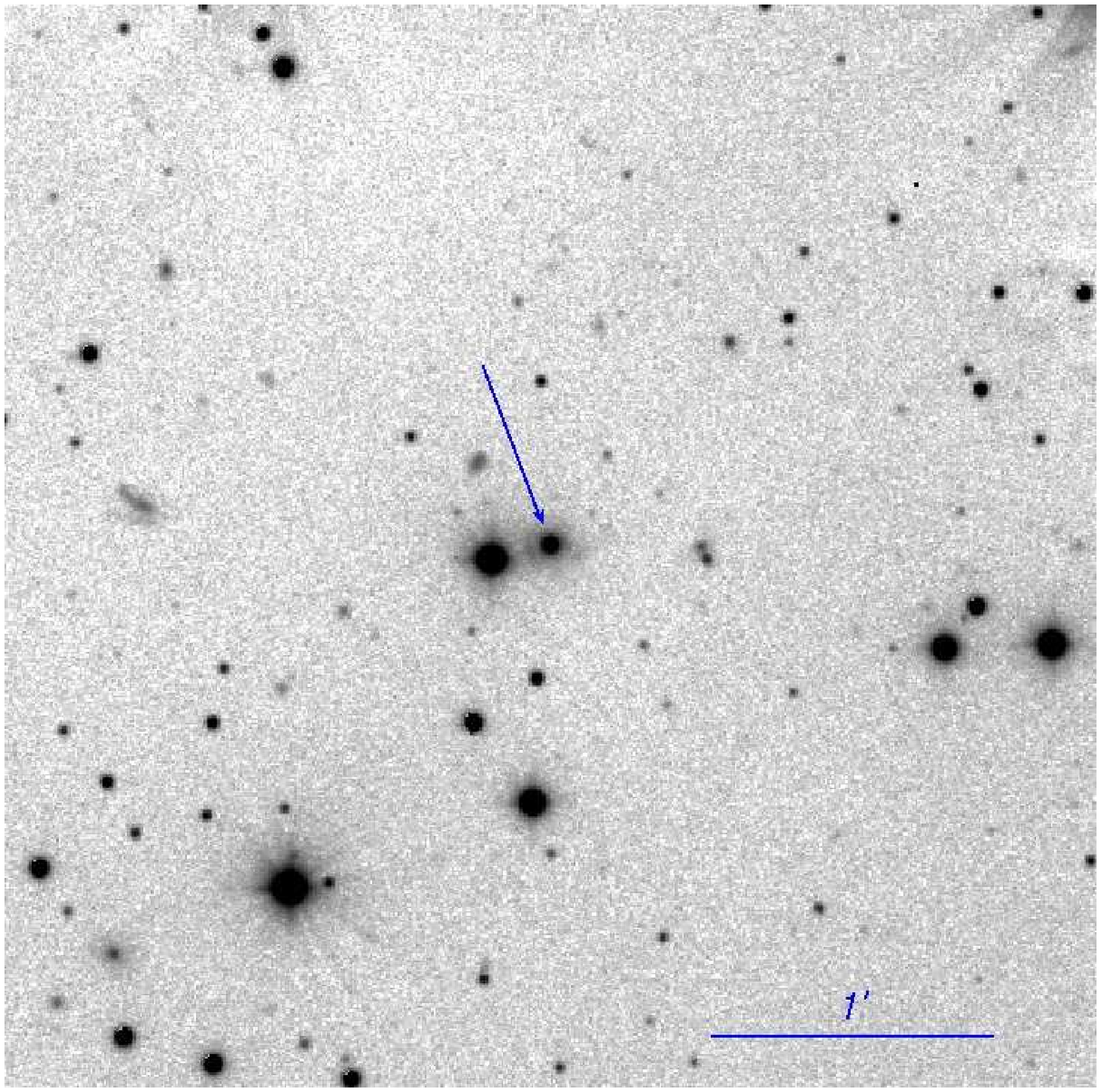} \\
  \end{tabular}
  \caption{--- Optical images of the fields of hard X-ray sources.}
  \label{fig:fcharts}
\end{figure*}

The finding charts, where the arrows show the optical objects identified
with hard X-ray sources, are shown in Fig.~\ref{fig:fcharts}. For the
sources IGR\,J01528$-$0326 and IGR\,J02343$+$3229 the arrow show the
position of the slit of the spectrometer. In the charts we show the direct
images in $r'$ filter (\emph{SDSS}), obtained with RTT150 telescope. In case
of X-ray source IGR\,J16562$-$3301, when we not able to obtain the direct
image of sufficient quality, the image taken from DSS-II red plate is shown.

\begin{figure*}
  \centering
  \begin{tabular}[c]{ll}
    \phantom{00000}IGR\,J01528$-$0326 &\phantom{00000}IGR\,J02343$+$3229 \\
    \plotspectrum{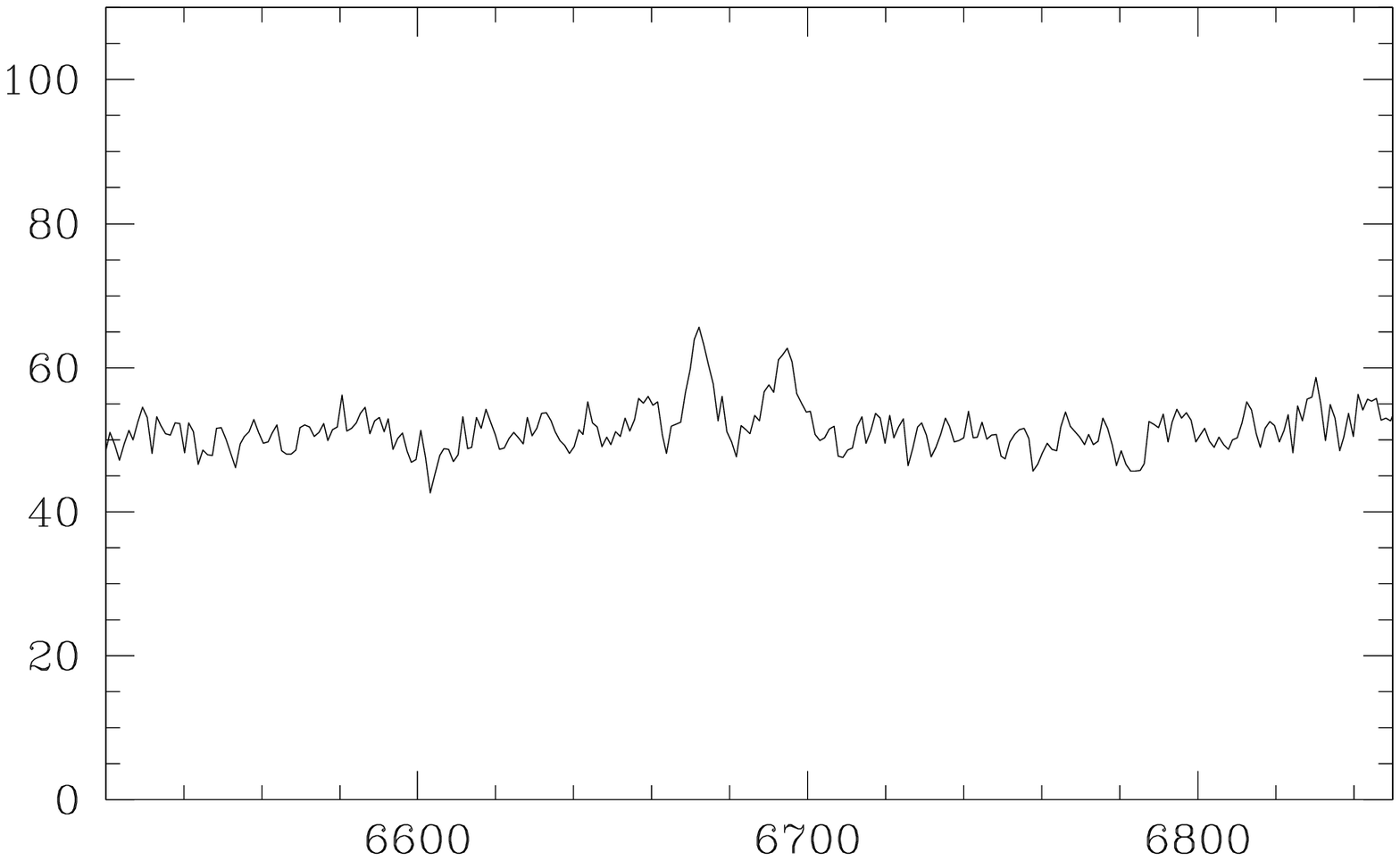}{$\lambda$, \AA}{$F_\lambda, \times
      10^{-16}$erg~s$^{-1}$~cm$^2$} & 
    \plotspectrum{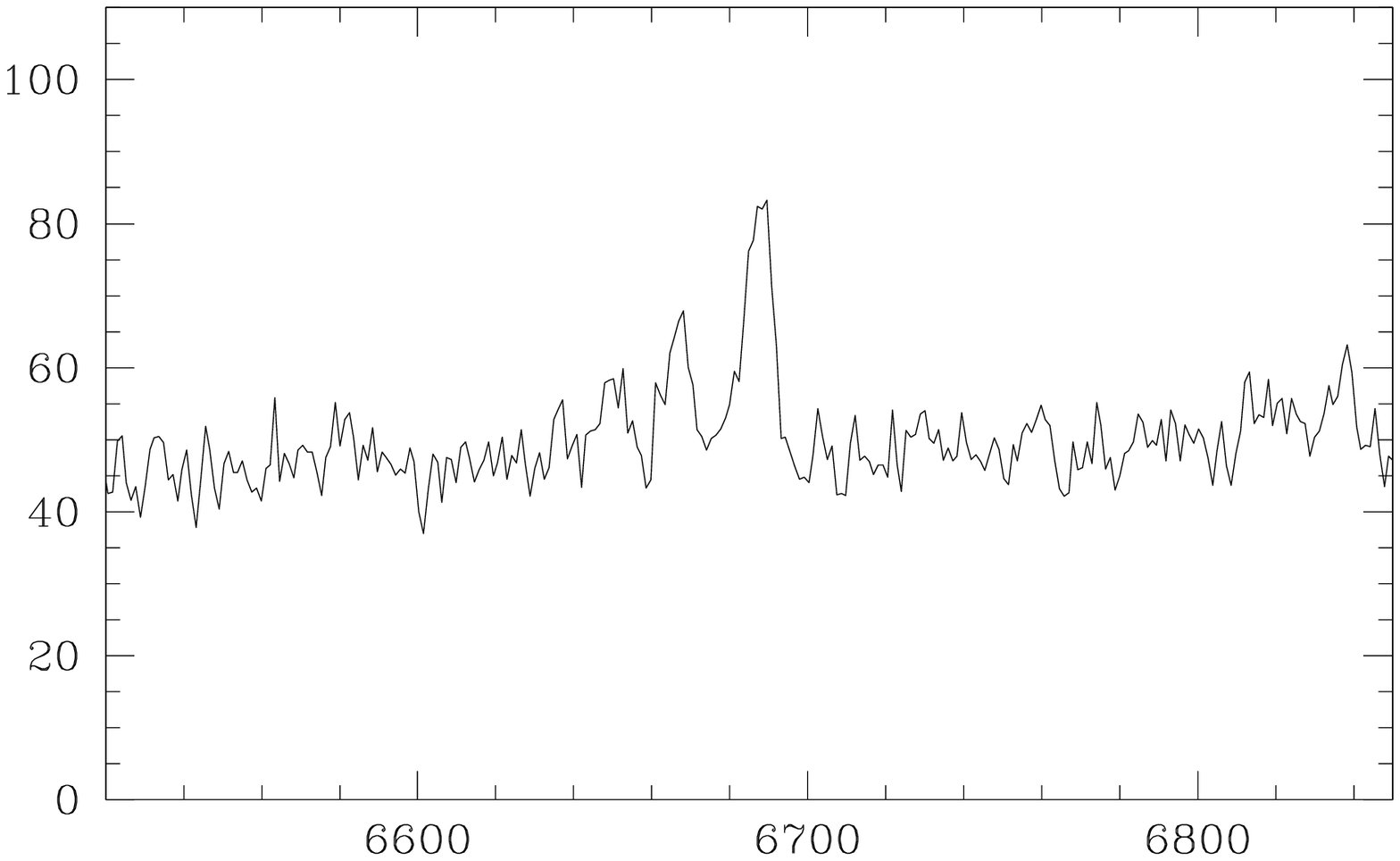}{$\lambda$, \AA}{$F_\lambda, \times
      10^{-16}$erg~s$^{-1}$~cm$^2$} \\
    \phantom{00000}IGRJ03334+3718 & \phantom{00000}IGRJ13038+5348\\
    \plotspectrum{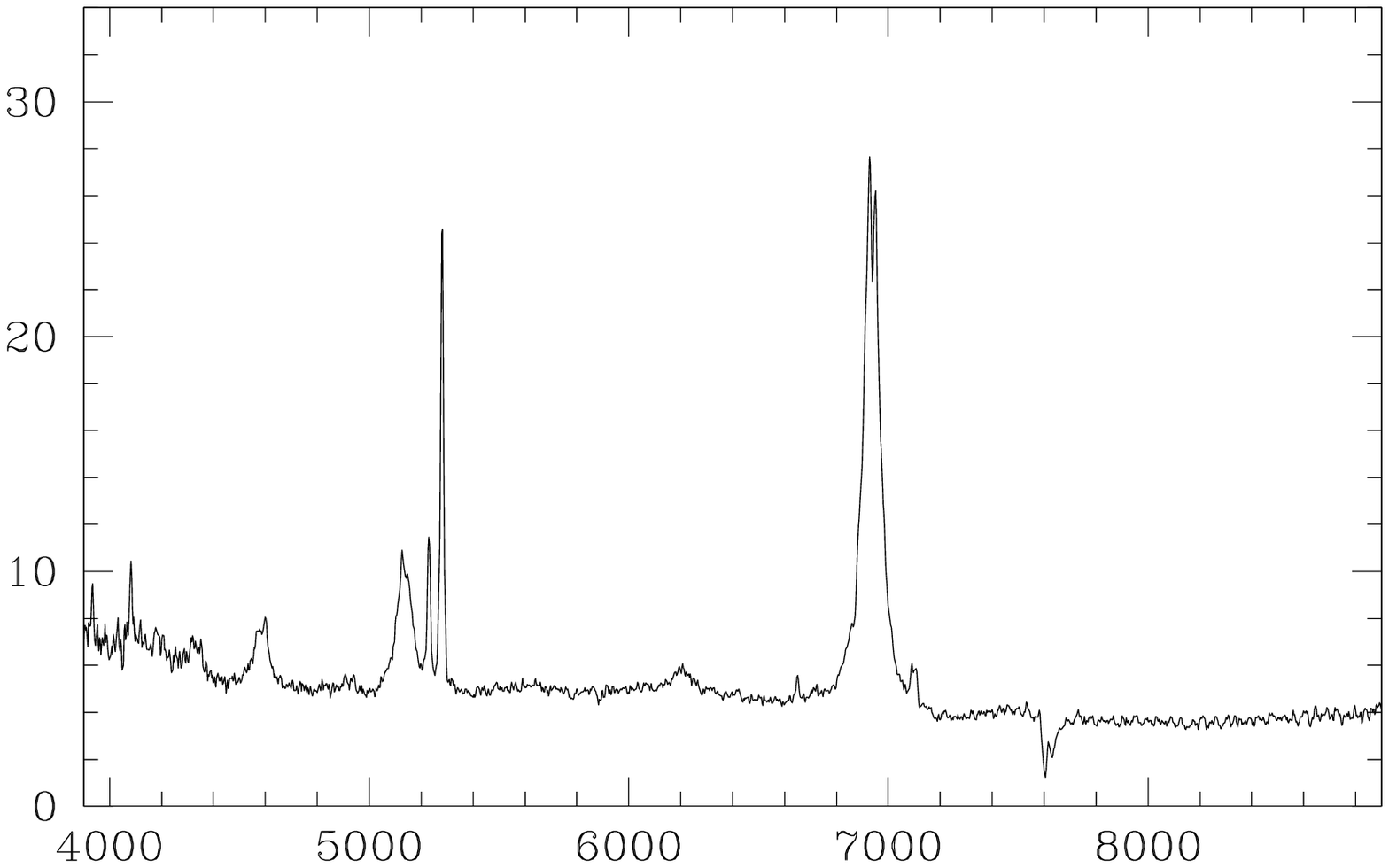}{$\lambda$, \AA}{$F_\lambda, \times
      10^{-16}$erg~s$^{-1}$~cm$^2$} & 
    \plotspectrum{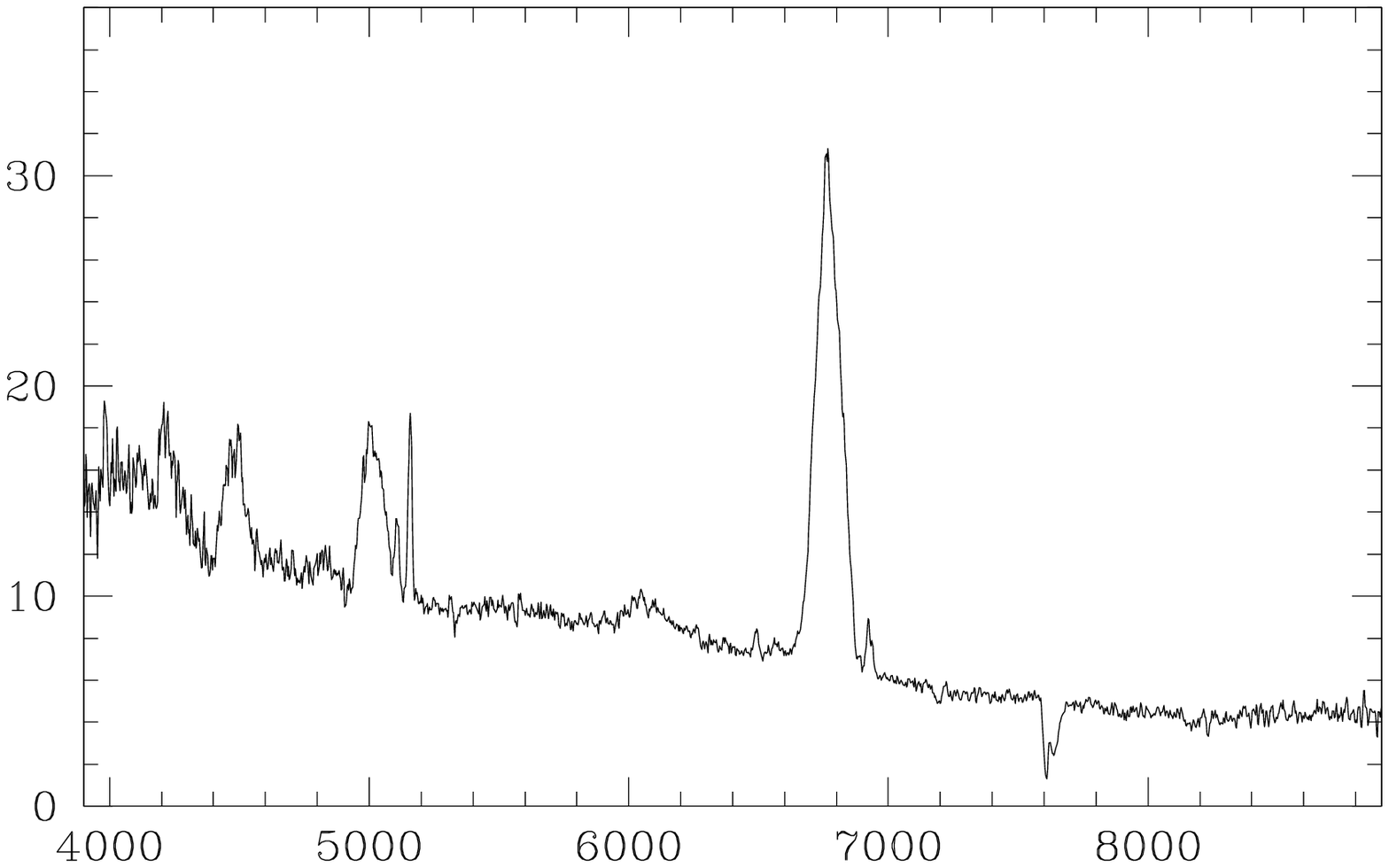}{$\lambda$, \AA}{$F_\lambda, \times
      10^{-16}$erg~s$^{-1}$~cm$^2$} \\
    \phantom{00000}IGR\,J16562$-$3301 &  \phantom{00000}SWIFT\,J1745.4$+$2906\\
    \plotspectrum{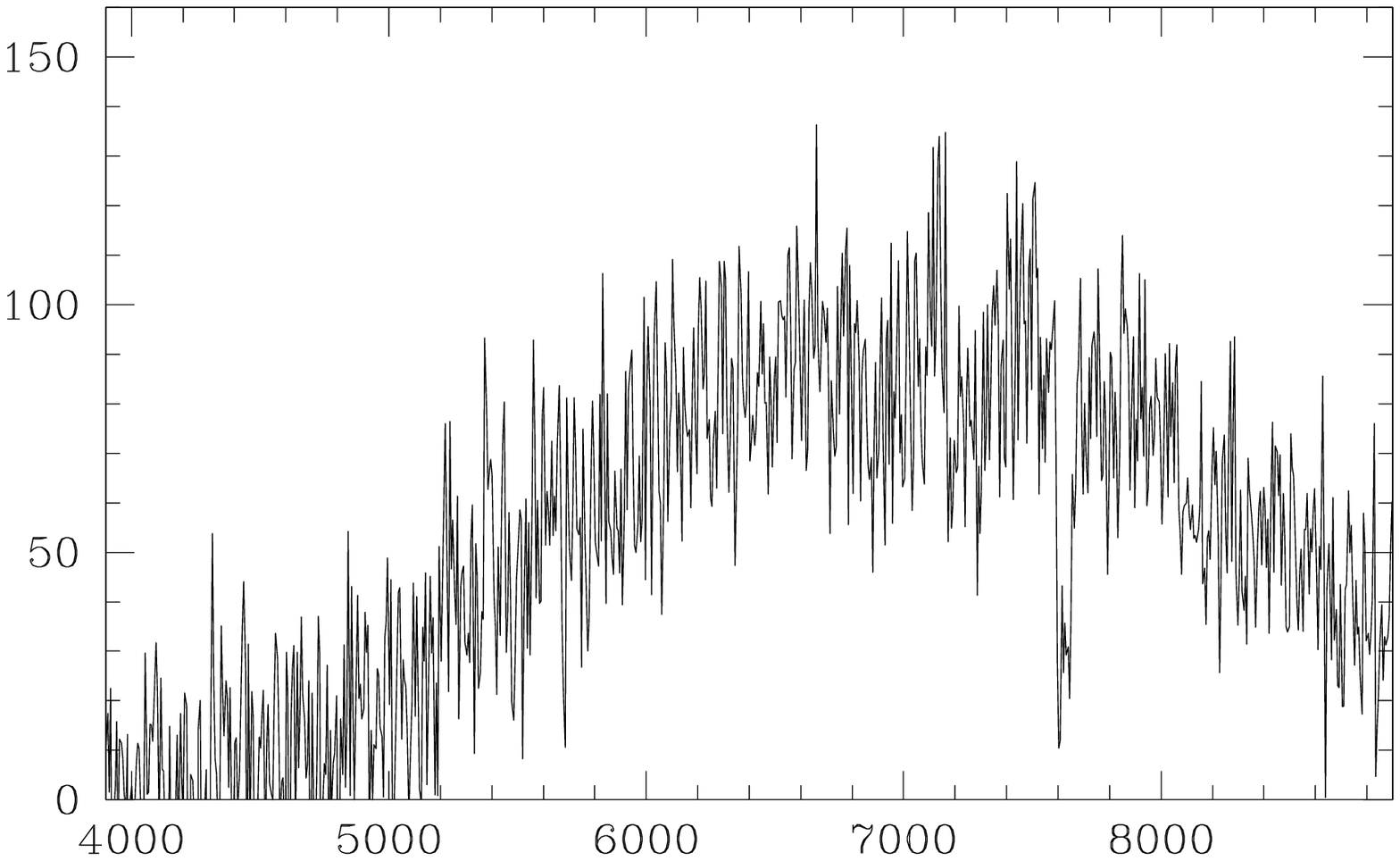}{$\lambda$, \AA}{отсчеты} &
    \plotspectrum{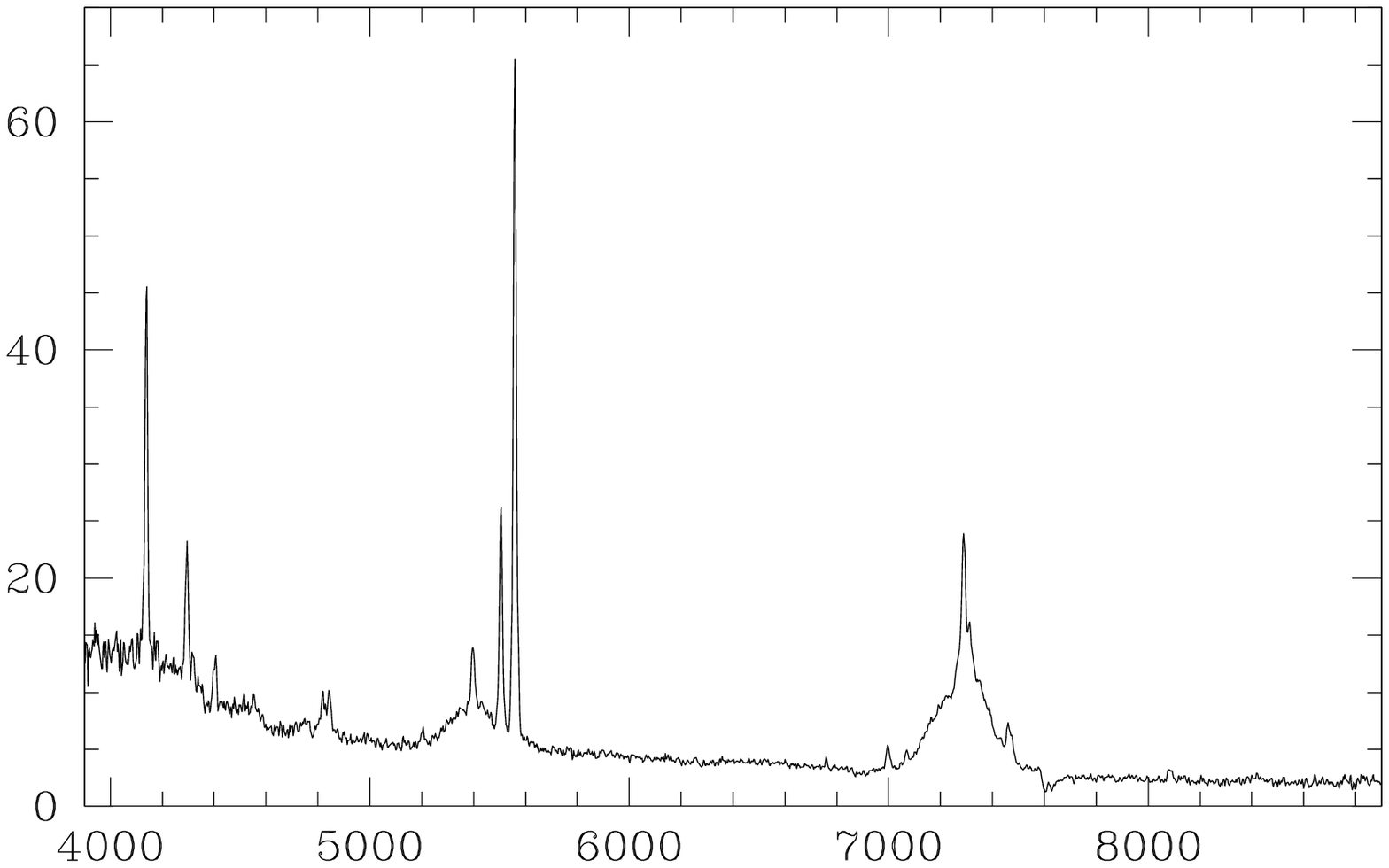}{$\lambda$, \AA}{$F_\lambda, \times
      10^{-16}$erg~s$^{-1}$~cm$^2$}\\
  \end{tabular}
\caption{--- Spectra of optical objects, which were identified with hard
  X-ray sources.}
  \label{fig:spectra}
\end{figure*}

Fig.~\ref{fig:spectra} shows spectra of optical objects, which were used to
identify the types of the objects. All spectra are converted to physical
fluxes, using the observations of spectrophotometric standard stars from the
list of \cite{massey88}. In cases of nearby galaxies MCG~$-$01$-$05$-$047
and NGC\,973 (X-ray sources IGR\,J01528$-$0326 and IGR\,J02343$+$3229), only
small fraction of the flux of the central parts of the galaxies fall into
the slit of the spectrometer. This produce the uncertainty of the
photometric calibration of the emission line flux, because the real angular
size of the line emitting region is unknown. For this objects we corrected
the flux for the slit size, assuming that the angular radius of the emitting
region is equal to 9\arcsec and 12\arcsec correspondingly, where the angular
sizes are taken so that we obtain the \emph{upper} estimate of the flux in
emission lines. In other cases the slit size correction was applied,
assuming that the source is point-like, which is a good approximation, since
all the other AGNs have the stellar-like nuclei.

Due to the uncertainty of photometrical conditions during our observations,
photometric calibration may contain significant systematic errors,
especially in blue part of the spectrum $\lambda<5500~$\AA. The spectra in
Fig.~\ref{fig:spectra} are not corrected for the Galactic interstellar
extinction. The spectrum of X-ray source IGR\,J16562$-$3301 is shown in CCD
counts, because it was observed at very large zenith angles and the overall
form of its spectrum is strongly distorted.

\begin{figure*}
  \centering
  \begin{tabular}[c]{ll}
    \large SWIFT\,J1930.5$+$3414 & \large 1RXS\,J193347.6$+$325422 \\
    \includegraphics[width=0.8\columnwidth]{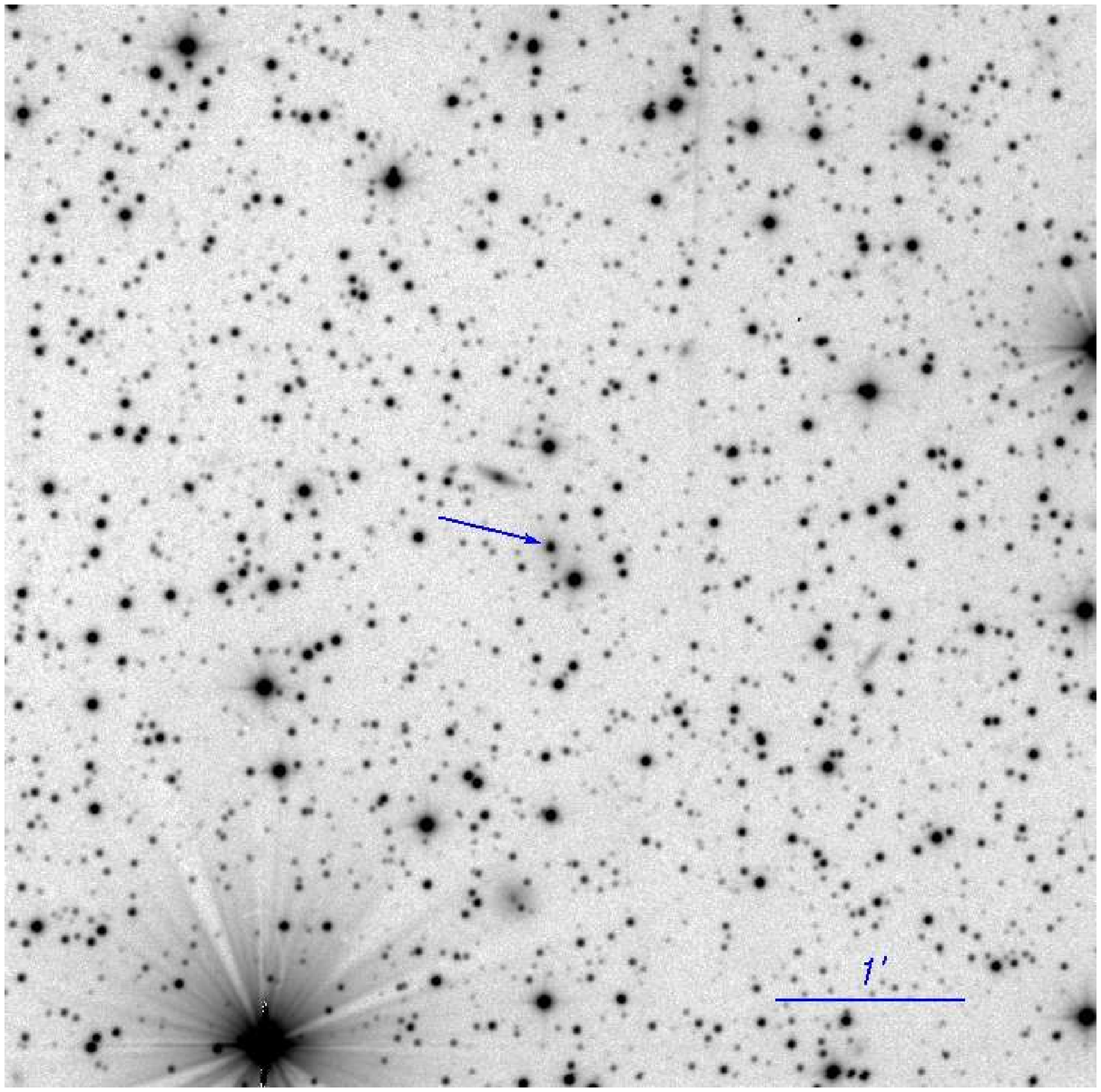} & 
    \includegraphics[width=0.8\columnwidth]{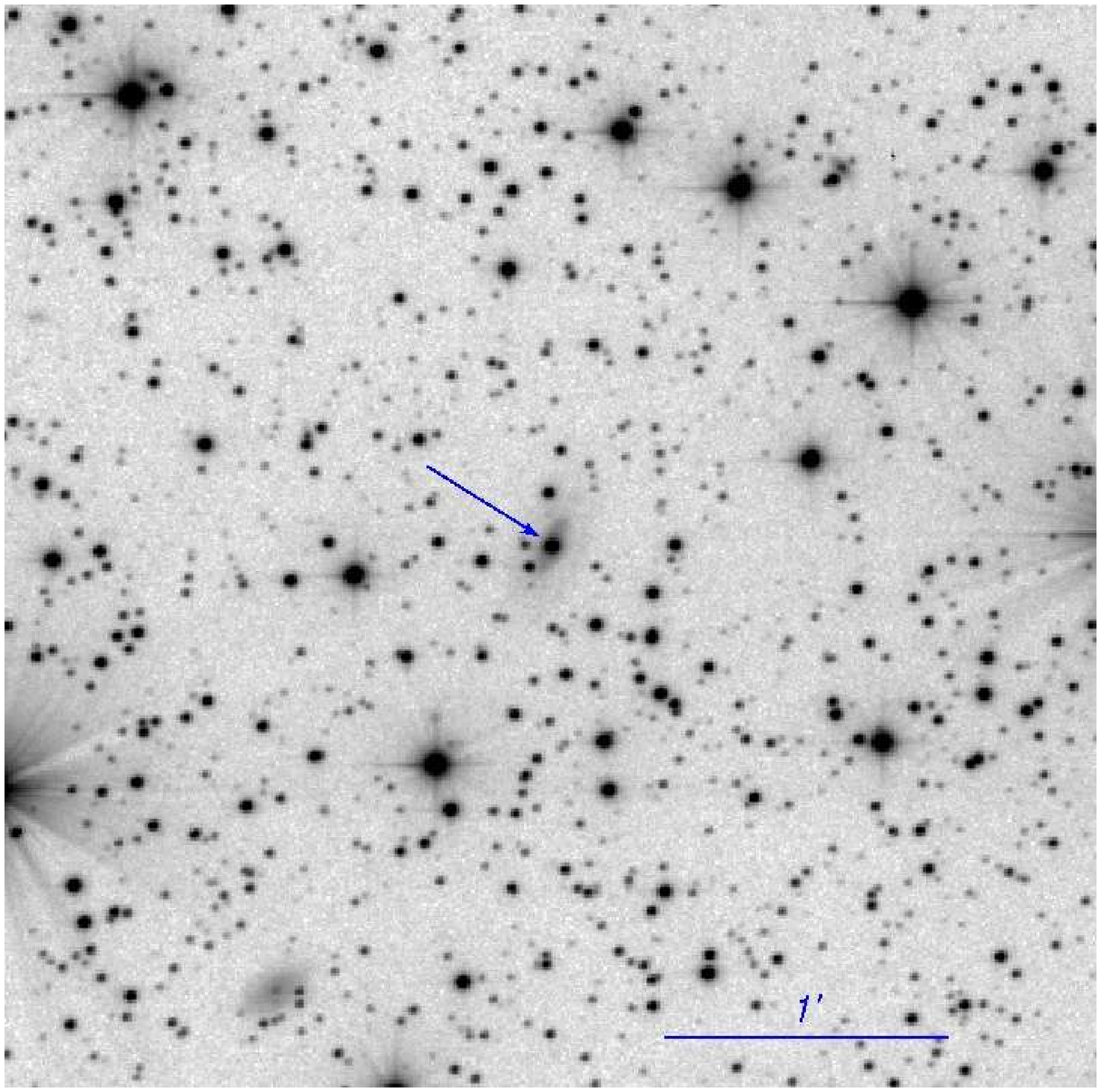} \\
    \\
    \large RX\,J2135.9$+$4728 \\ 
    \includegraphics[width=0.8\columnwidth]{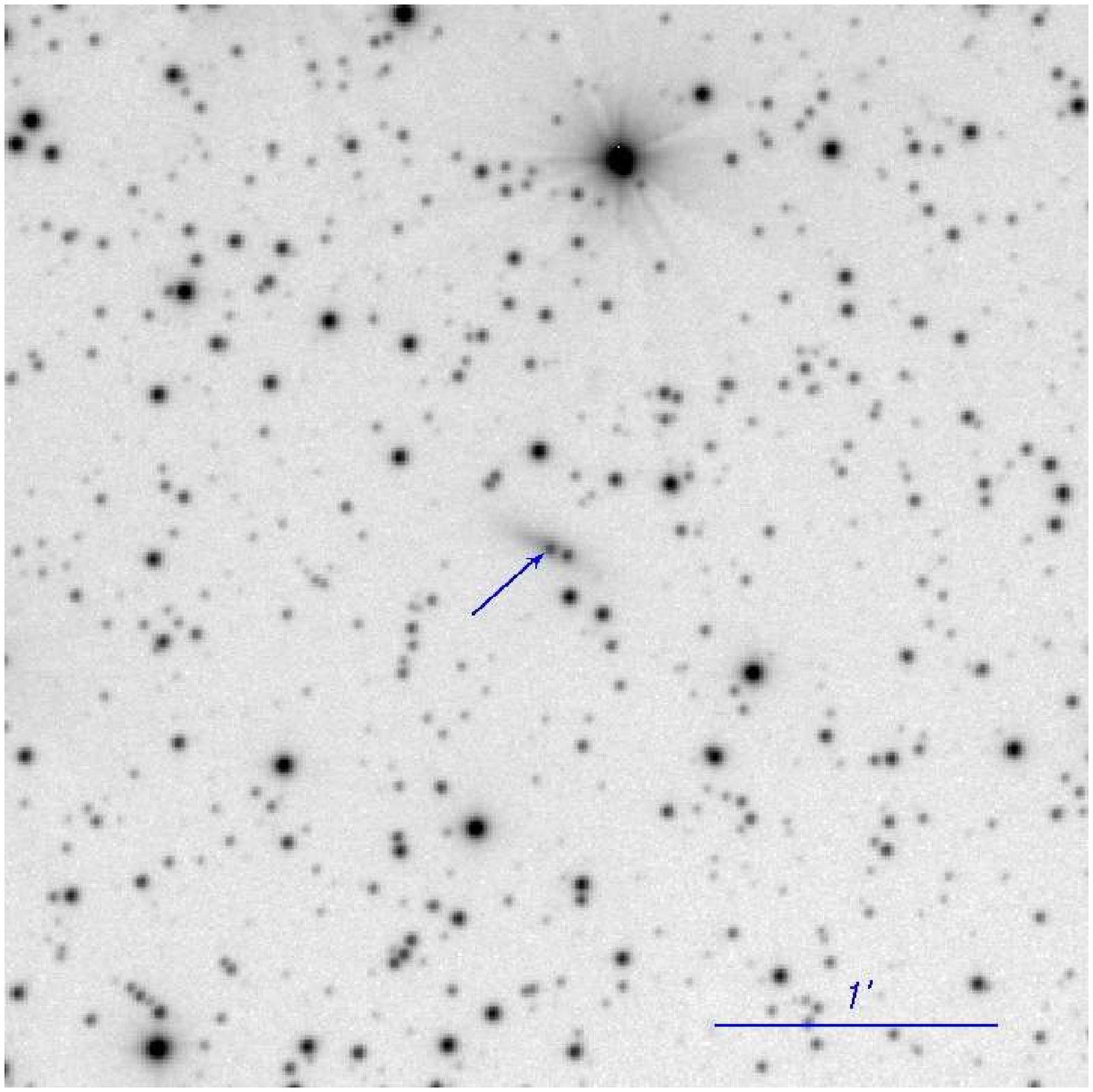} \\
  \end{tabular}
  \bigskip

   {\small\textbf{Fig.~\ref{fig:fcharts}.} --- \emph{Continued}.}
\end{figure*}

\begin{figure*}
  \centering
  \begin{tabular}[c]{ll}
     \phantom{00000}SWIFT\,J1930.5$+$3414 & \phantom{00000}1RXS\,J193347.6$+$325422\\
    \plotspectrum{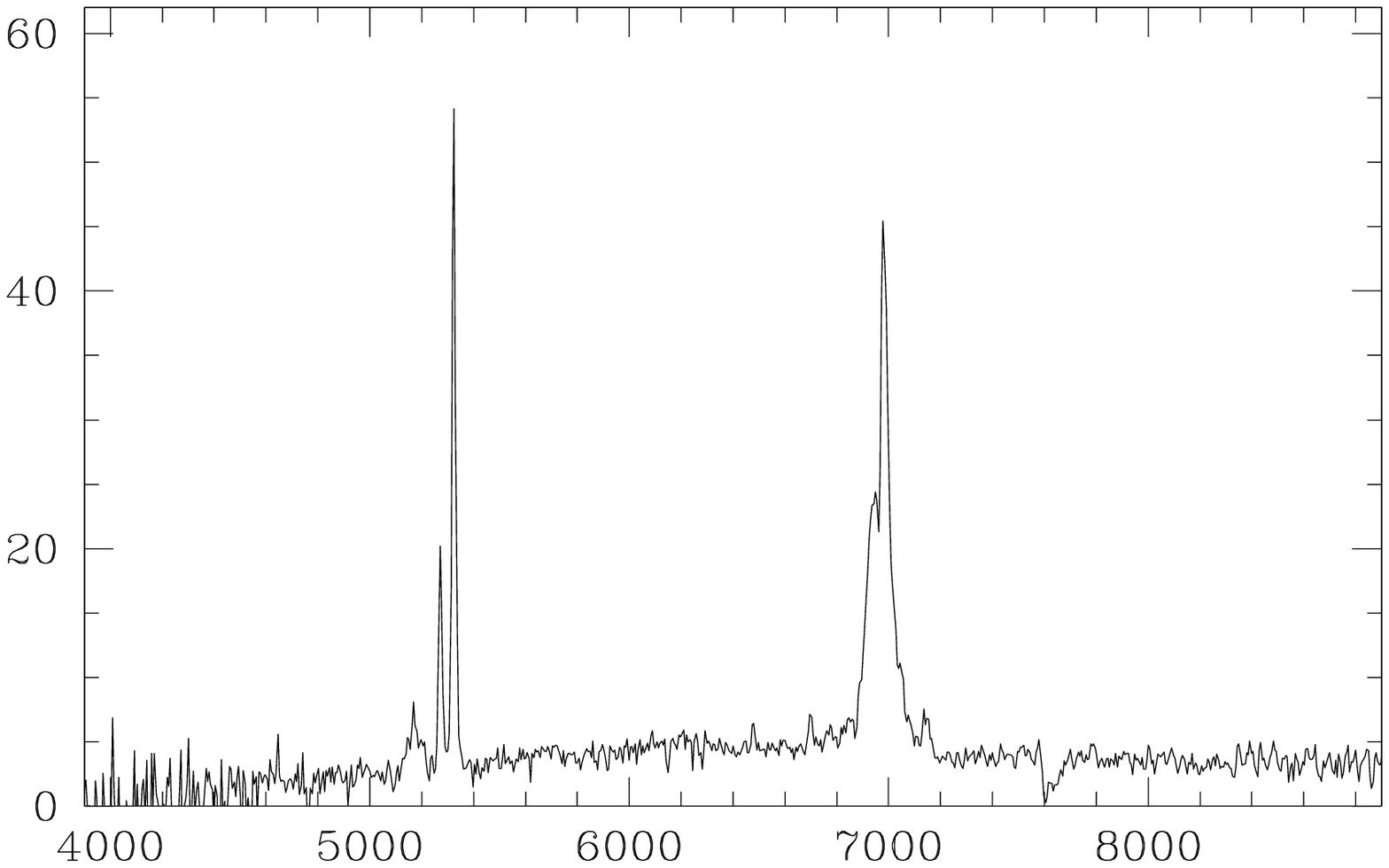}{$\lambda$, \AA}{$F_\lambda, \times
      10^{-16}$erg~s$^{-1}$~cm$^2$} &
    \plotspectrum{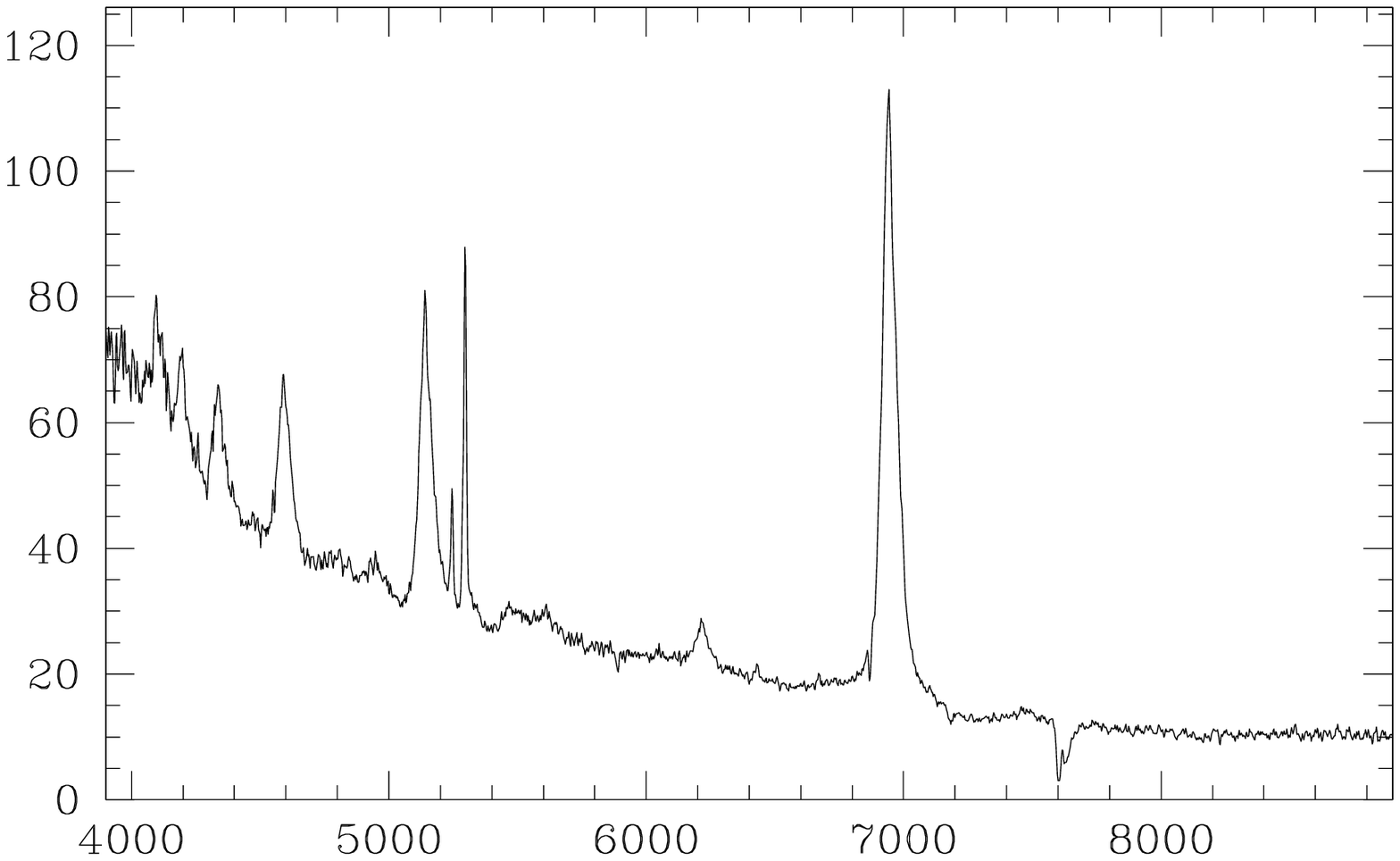}{$\lambda$, \AA}{$F_\lambda, \times
      10^{-16}$erg~s$^{-1}$~cm$^2$}\\
    \phantom{00000}RX\,J2135.9$+$4728\\
    \plotspectrum{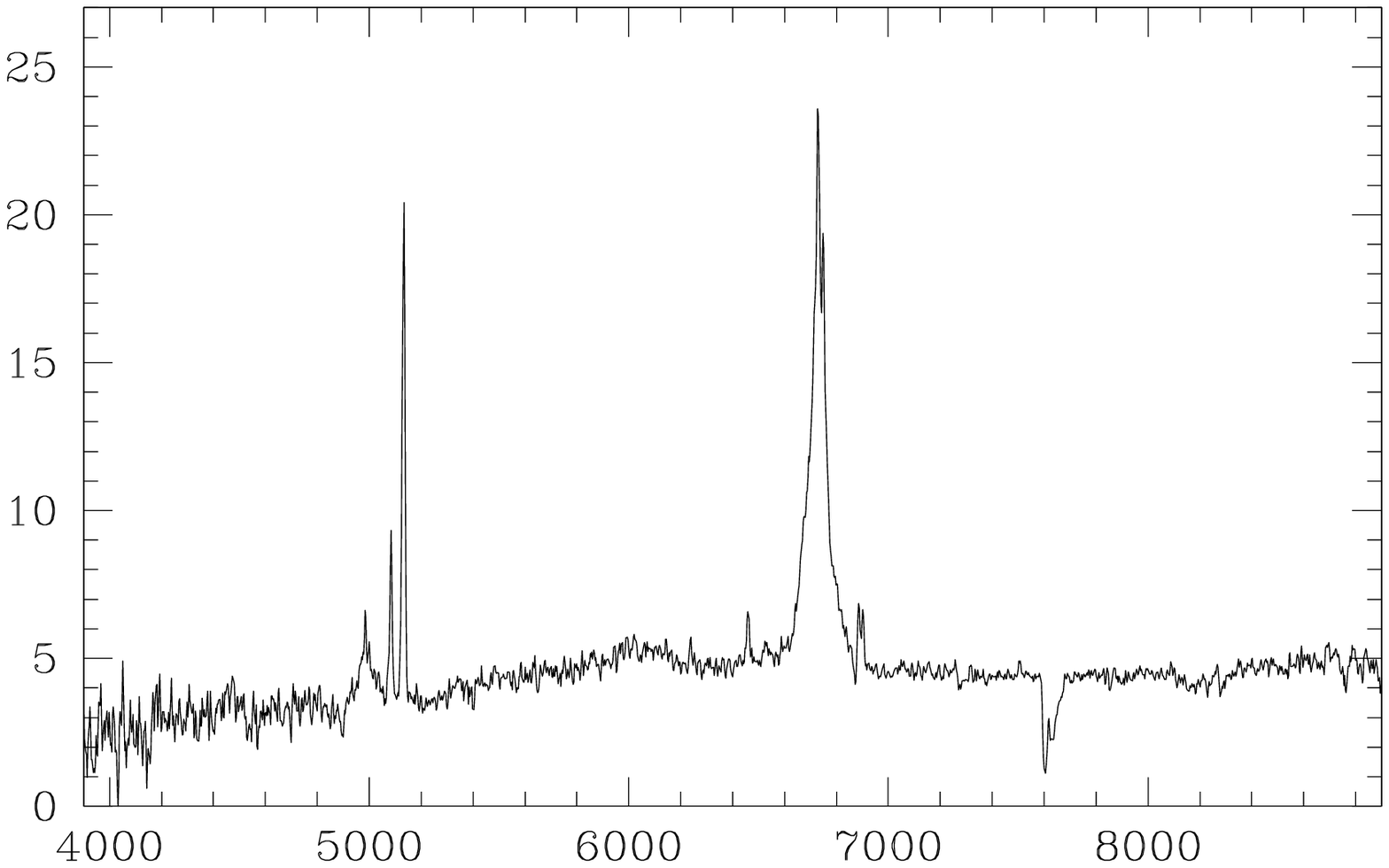}{$\lambda$, \AA}{$F_\lambda, \times
      10^{-16}$erg~s$^{-1}$~cm$^2$} \\
  \end{tabular}
  \bigskip

  {\small\textbf{Fig.~\ref{fig:spectra}.} --- \emph{Continued}.}
\end{figure*}

\subsection{Notes on individual objects}
\label{sec:notes}

IGR\,J01528$-$0326 --- The source is unambiguously identified with nearby
edge-on spiral galaxy MCG~$-$01$-$05$-$047. Optical spectrum of the central
part of the galaxy shows narrow emission lines $H_\alpha$, [NII],6583 and
[SII],6716. From the intensity ratio $\lg(\mathrm{[NII],6583}/H_\alpha)
\approx 0$, we conclude that this source most probably is a Seyfert 2 AGN
\citep{veilleux87}.

IGR\,J02343$+$3229 --- The source is identified with nearby edge-on spiral
galaxy NGC973. Optical spectrum of the central part of the galaxy shows
narrow emission lines \oiii,5007, $H_\alpha$, [NII],6583 and [SII],6716. The
emission lines intensity ratios can be estimated as
$\lg(\mathrm{\oiii,5007}/H_\beta) > 0.7$ and
$\lg(\mathrm{[NII],6583}/H_\alpha) \approx 0.4$. Therefore, this source is
most probably a Seyfert 2 AGN \citep{veilleux87}.

IGR\,J03334$+$3718 --- The source was detected also in \emph{ROSAT} All-Sky
Survey (1RXS\,J033316.2$+$371815), which allow to measure more accurate
position with the error circle $\approx10\arcsec$. After that this source is
unambiguously identified with the galaxy with stellar-like nucleus. The
optical spectrum of the nucleus contains redshifted emission lines --- broad
Balmer lines, and also narrow lines \oiii\ 4959,5007 and others, which allow
to identify it as Seyfert 1 AGN.

IGR\,J13038$+$5348 --- This source was observed by SWIFT XRT. Using these
data, we determined the source position with the few arcseconds accuracy,
which allow to associate this source with bright nearby galaxy MCG
$+$09$-$21$-$096. It was previously detected as an emission line galaxy in
Second Burakan Spectral Survey \citep{markaryan8385,bicay00}. The spectrum
of the central part of this galaxy contain broad Balmer and narrow emission
lines, identifying this object as a Seyfert 1 AGN.

IGR\,J16562$-$3301 --- The accurate position of hard X-ray source was
obtained using SWIFT XRT data \citep{tueller06}. Using the same X-ray data
we determine even more accurate position of this source. Systematic
uncertainty of XRT aspect solution was strongly reduced using two other
X-ray sources which were identified with two relatively bright stars in
optical band \citep{atel1270}. In this way we obtain the following X-ray
source coordinates $\alpha, \delta$: $16$:$56$:$16.82$, $-33$:$02$:$12.5$
(J2000, here the astrometry was done relative to \emph{UCAC2} catalog). The
positional error here should be almost purely statistical and is about
$1.5\arcsec$.

These coordinates agree with the position of very bright radio source
NVSS\,J165616-330211 (the flux is about 0.4~Jy at 1.4~GHz). The coordinates
of the only optical object, which is coincide with both X-ray and radio
sources, are given in Table~\ref{tab:src}. In the optical spectrum of this
object there are no bright emission lines (Fig.~\ref{fig:spectra}). All
these data allow to conclude, that this source is a BL Lacertae object. In
order to measure its redshift, the spectrum with much higher signal-to-noise
ratio is needed.

\mbox{SWIFT\,J1745.4$+$2906, SWIFT\,J1930.5$+$3414,} \mbox{ }
1RXSJ193347.6$+$325422 --- These hard X-ray sources were detected by SWIFT
observatory. The SWIFT XRT data, available in public domain, allowed to
unambiguously identify these objects with the galactic nuclei. The optical
spectra, obtained with RTT150 telescope, show that all of them are Seyfert 1
AGN (Fig.~\ref{fig:spectra}).

RX\,J2135.9$+$472 --- As Seyfert 1 AGN this object was identified earlier,
in frames of the work on optical identifications of X-ray sources from
\emph{ROSAT} All-Sky Survey \citep{motch97}. However, this object was not
included in \emph{NED} database, and we identified this object
independently. In Table~\ref{tab:src} we provide more accurate redshift, as
compared to that published by \cite{motch97}. Apart of the galactic nucleus,
there is another one stellar like object in the image of the galaxy (see
Fig.~\ref{fig:fcharts}). We obtained the spectrum of this source and show
that this is a chance projection of a star.

\begin{table}
  \centering
  \caption{The AGN luminosities in X-ray 17--60~keV band and in
    \mbox{[OIII],5007} emission line.}
  \label{tab:lum}
  \medskip
  \renewcommand{\tabcolsep}{0.35cm}
  \begin{tabular}{lccclcllccc}
    \hline
    \hline
    Name &  $\lg L_{X}$& \phantom{$<$}$\lg L_{[OIII]}$ \\
    \hline
    IGR\,J01528$-$0326       & 43.02 &          $<$39.08  \\    
    IGR\,J02343$+$3229       & 43.35 & \phantom{$<$}40.16 \\    
    IGR\,J03334$+$3718       & 44.15 & \phantom{$<$}42.05 \\    
    IGR\,J13038$+$5348       & 43.80 & \phantom{$<$}40.56 \\    
    SWIFT\,J1745.4$+$2906    & 44.71 & \phantom{$<$}42.50 \\    
    SWIFT\,J1930.5$+$3414    & 44.12 & \phantom{$<$}42.21 \\    
    1RXS\,J193347.6$+$325422 & ---   & \phantom{$<$}42.13 \\    
    RX\,J2135.9$+$4728    & 43.24 & \phantom{$<$}41.04 \\    
    \hline 
  \end{tabular}
  
  \medskip
  \begin{minipage}{1.0\linewidth}
  \end{minipage}
\end{table}


\section{Discussion}

In Table~\ref{tab:lum} we give the luminosities in hard X-rays (17--60~keV)
and in \oiii,5007 emission line\footnote{To convert fluxes into the
  luminosities we use the cosmological model with $\Omega=0.3$,
  $\Lambda=0.7$, $H_0=71$~km~s$^{-1}$~Mpc$^{-1}$.}. The fluxes in X-rays are
taken from the catalogs of INTEGRAL and SWIFT all-sky surveys
\citep{krivonos07,tueller07}. In last case the fluxes were converted to
17--60~keV band assuming power law spectrum with photon index $\Gamma=1.8$.
There is no X-ray flux for the source 1RXS\,J193347.6$+$325422 in
Table~\ref{tab:lum}, because this source is not included in current version
of SWIFT survey catalog. The fluxes in \oiii,5007 emission line are
corrected for the Galactic interstellar extinction using data from
\cite{schlegel98}.

\begin{figure}
  \centering \smfigurewocap{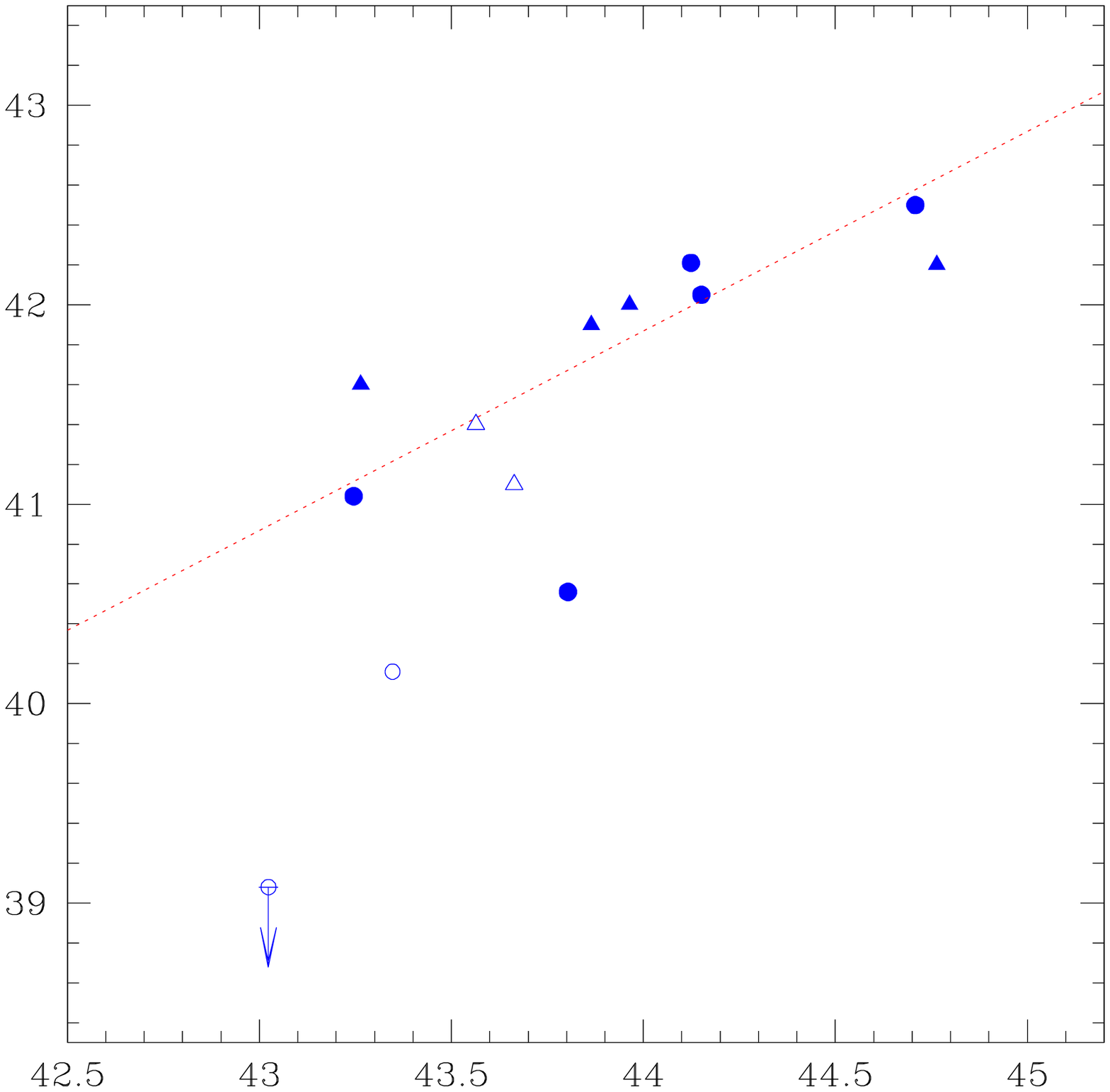}{$\lg L_{X,17-60\,\mbox{\footnotesize
        keV}}$}{$\lg L_{[OIII],5007}$}
  \caption{The relation between the luminosities in X-ray 17--60~keV band
    and in \oiii,5007 emission line. Filled and open symbols show Seyfert
    galaxies of type 1 and 2 respectively. Circles show the data from our
    present work, triangles --- the data which we obtained earlier
    \citep{bikmaev2006a}. Dashed line show the dependency $\lg
    L_X/L_{[OIII]} = 2.13$ (see text).}
  \label{fig:lum}
\end{figure}

In Fig.~\ref{fig:lum} we show the relation between the luminosities in
17--60~keV X-ray band and in \mbox{\oiii,5007} line. Here we also add the
data for AGNs, which we optically identified earlier \citep{bikmaev2006a}.
The data for the main part of AGNs are well agree with the known correlation
\citep[see, e.g.,][]{heckman05}. Excluding three lower points in
Fig.~\ref{fig:lum}, we find that the mean and the dispersion of the
luminosities ratio logarithm are equal $2.13\pm0.08$ and $0.27$
respectively, in good agreement with the results for AGNs, selected mainly
in optical band \citep{heckman05}.

In contrast to optically selected AGNs, our sample contain AGNs with much
lower luminosity \oiii\ line, as should be expected for their X-ray
luminosities. First, these are two nearby AGNs in MCG~$-$01$-$05$-$047 and
NGC\,973, nearly exactly edge-on galaxies. As it was noted above, in the
calibrations of their spectra there is a considerable uncertainty, due to
the unknown angular size of the line emitting region. However, their
spectra were calibrated so that to obtain an \emph{upper} estimate of \oiii\
line fluxes (see above) and the lack of luminosity in this line should exist
anyway. Most probably, this lack of luminosity in \oiii\ line is explained
by the extinction in the disks of the galaxies, since in these two cases the
galaxies are turned nearly exactly edge-on. From the observed thickness of
the disk, we estimate that the size of absorbed region should not be larger
than $\approx$1~kpc.

In Fig.~\ref{fig:lum} the point, correspondent to the AGN in hard X-ray
source IGR\,J13038$+$5348, is also shifted down. In the optical image of
this source (Fig.~\ref{fig:fcharts}) on can see that here the galaxy is also
approximately edge-on. However, in the optical spectrum of this object there
are broad Balmer lines. Probably, the other mechanisms of the weakening of
narrow forbidden lines are in effect in this case. 

In their work \cite{heckman05} conclude, that AGN samples, selected using
\oiii\ emission line flux are as complete as those selected in hard X-rays.
In our work we are faced on with few cases, where the logarithm of the
\oiii\ line and hard X-ray luminosities ratio was significantly lower than
$\approx2.1$. These objects should modify the selection of AGN using their
flux in \oiii\ line and change the form of the AGN luminosity function in
\oiii\ line as compared to that in hard X-rays. The fraction of these AGNs
may be less than $\sim20\%$ (as it follows from Fig.~\ref{fig:lum}), since
here we consider only, probably, most absorbed AGNs, which were not known
earlier and were detected only using observations in hard X-rays.

\section{Conclusions}

In this paper we present our new results of the observations of the fields
of unidentified hard X-ray sources, detected in statistically complete
all-sky surveys. The observations were carried out using Russian-Turkish
1.5-m Telescope (RTT150). Nine X-ray sources were identified with the active
galactic nuclei. Two of them are hosted by nearby, nearly exactly edge-on,
spiral galaxies MCG~$-$01$-$05$-$047 and NGC\,973. One source,
IGR\,J16562$-$3301, is most probably BL Lac object (blazar). Other AGNs are
observed as stellar-like nuclei of spiral galaxies, with broad emission lines
in their spectra. All of them, excluding probably IGR\,J16562$-$3301, are
nearby objects --- their redshifts are $z<0.1$, as for the majority of AGNs
in INTEGRAL All-Sky Survey. Almost all of them are located at high Galactic
latitudes $|b|>5^\circ$.

Taking in account the data, which we obtained earlier \citep{bikmaev2006a},
we show, that the luminosities in hard X-rays (17--60~keV) and in \oiii,5007
emission line, $\lg L_X/L_{[OIII]} \approx 2.1$, are well correlated, as
expected, for the majority of AGNs selected in hard X-rays. However, few
AGNs considerably deviate from this correlation. For example, the flux in
\oiii\ line turns to be lower in AGNs in two nearby edge-on spiral galaxies
MCG~$-$01$-$05$-$047 and NGC\,973, which can be explained by the extinction
in their galactic disk. The fraction of AGNs, for which the \oiii\ line
fluxes are significantly deviate from that correlation, may be about $20\%$,
which should significantly modify \oiii\ line AGN selection and the form of
the AGN luminosity function in \oiii\ line as compared to that in hard
X-rays. 

In result of the recent works on optical identifications of hard X-ray
sources \citep[e.g.,][and also this
work]{bikmaev2006a,bikmaev2006b,masetti06}, the catalog of INTEGRAL All-Sky
Survey \citep{krivonos07} is now almost complete at high Galactic latitudes.
Now there are only few X-ray sources, which remain optically unidentified,
in this catalog at Galactic latitudes $|b|>5^\circ$. At the same time, only
in subsample of AGNs from this survey, 33 out of 127 objects were unknown
before the survey was started \citep{sazonov07}.

At low Galactic latitudes there are few tens of hard X-ray sources from
INTEGRAL survey, which are not optically identified yet. We will continue
our works on their optical identifications. Optical objects, associated with
some of these objects turn to be too faint for the observations with RTT150.
Now we submitted the proposals to observe these fainter objects with larger
telescopes.

\acknowledgements

This work is supported by grants RFFI 05-02-16540, RFFI 07-02-01004,
NSh-784.2006.2 and NSh-1100.2006.2, and also with programs of Russian
academy of sciences P-04 and OFN-17.
\medskip

Note: When this paper was already accepted for publication we have noticed
that one of our hard X-ray sources IGR\,J16562$-$3301 was independently
identified as BL Lacertae object by \cite{masetti08}, who also measured its
redshift $z=2.40$.

\end{document}